\documentclass[smallextended]{svjour3}       
\smartqed  

\usepackage{amsmath}
\usepackage{algorithm}
\usepackage{algorithmic}
\usepackage{parselines}
\usepackage{epsfig}
\usepackage[nottoc]{tocbibind}
\usepackage[colorlinks,citecolor=blue,urlcolor=blue,filecolor=blue,backref=page]{hyperref}
\usepackage{graphicx}
\usepackage{graphics}
\usepackage{subfigure}
\usepackage{verbatim}
\usepackage{float}
\usepackage{mathtools}

\journalname{}

\begin{document}

\title{Hierarchical EM algorithm for estimating the parameters of Mixture of Bivariate Generalized Exponential distributions} 


\titlerunning{MBVGE}        
\author{Arabin Kumar Dey \\ Debasis Kundu \\ Tumati Kiran Kumar
}

           

\institute{A. K. Dey \at
              Department of Mathematics \\
              IIT Guwahati\\
              Guwahati\\
              Assam\\
              Tel.: +91361-258-4620\\
              \email{arabin@iitg.ac.in}           
             \and
             D. Kundu \at
             Department of Mathematics and Statistics,\\ 
             IIT Kanpur,\\   
             Kanpur, India\\
             \email{kundu@iitk.ac.in} \\
             \and
             Tumati Kiran Kumar \at
             Amazon India,\\ 
             Hyderabad,\\   
              \email{classykiran@gmail.com} \\
}






\maketitle

\begin{abstract}

 This paper provides a mixture modeling framework using the bivariate generalized exponential distribution.  We study different properties of this mixture distribution. Hierarchical EM algorithm is developed for finding the estimates of the parameters. The algorithm takes very large sample size to work as it contains many stages of approximation. Numerical Results are provided for more illustration. 

\keywords{Joint probability density function; Bivariate Generalized Exponential distribution; Mixture distribution; Pseudo likelihood function; EM algorithm}
\end{abstract}

\section{Introduction}
\label{intro}

 In this paper we study mixture of two bivariate generalized exponential distributions.  We choose Marshall-Olkin type of bivariate generalized exponential distribution introduced by Gupta and Kundu \cite{GuptaKundu:2009} for this purpose.  The distribution can be used to model a data set which is heterogeneous and non-negative in nature where some of components are equal.  The main objective of this paper is to explore the issues related to estimation of the parameters for this bivariate mixture distribution through EM algorithm.  We see the behavior of EM algorithm over different sample size and parameters.  The calculation of the E and M step is little cumbersome.  An estimation procedure through hierarchical EM algorithm helps us to provide a computationally efficient procedure to get the parameter values.  The simulation study shows that the method works well mainly for large sample data.  It fails to provide the proper estimate when sample size is not sufficiently large.          
 
 Mixture distribution plays an important role in modeling heterogeneous populations, see for example McLachlan and Peel \cite{MclachlanPeel:2000}.  A Mixture distribution can easily capture Multimodality.  We can also bring the heavy tail behaviour by mixing two distributions.  An extensive work has been done on a mixture of multivariate normal distributions, not much work has been done on a mixture of multivariate non-normal distributions.  Recently mixture of bivariate Birnbaum Saunder distribution is introduced by Khosravi, Kundu and Jamalizadeh \cite{KhosraviKunduJamalizadeh:2015} to model the fatigue failure caused by cyclic loading.  For some related work in this connection readers are referred to \cite{SarhanBalakrishnan:2007}, \cite{BalakrishnanGuptaKunduLeivaSanhueza:2011}.  Mixture of bivariate generalized exponential is not used so far to model mixture of bivariate life time data.  It can be a good option to model such data sets.      
  
 The rest of the paper is organized as follows. In section 2, we provide the formulation of MBVGE distribution.  Some important properties for MBVGE are stated in section 3.  EM algorithm to compute the MLEs of the unknown parameters is provided in section 4. Discussion regarding Numerical Simulations and results are kept at Section 5. Finally we conclude the paper in section 6.

\section{Formulation of MBVGE}

  The univariate Generalized Exponential (GE) distribution has the following cumulative density function (CDF) and probability density function (PDF) respectively for $x > 0$;
\begin{eqnarray*}
F_{GE}(x; \alpha, \lambda) = (1 - e^{-\lambda x})^{\alpha}, ~~~~~  f_{GE}(x; \alpha, \lambda) =  \alpha \lambda e^{-\lambda x} (1 - e^{-\lambda x})^{\alpha - 1}
\end{eqnarray*} 
  Here $\alpha > 0$ and $\lambda > 0$ are shape parameter and scale parameters.  It is clear that for
$\alpha = 1$, it coincides with the exponential distribution. From now on a GE distribution with
the shape parameter $\alpha$ and the scale parameter $\lambda$ will be denoted by GE($\alpha$, $\lambda$). For brevity
when $\lambda = 1$, we will denote it by GE($\alpha$) and for $\alpha = 1$, it will be denoted by Exp($\lambda$).
From now on unless otherwise mentioned, it is assumed that $\alpha_1 > 0$, $\alpha_2 > 0$, $\alpha_3 > 0$, $\lambda > 0$.

Suppose $U_1 \sim GE(\alpha_1 , \lambda)$, $U_2 \sim GE(\alpha_2 , \lambda)$ and $U_3 \sim GE(\alpha_3 , \lambda)$ and they are mutually
independent. Here `$\sim$' means follows or has the distribution. Now define $X_1 = \max\{U_1 , U_3 \}$
and $ X_2 = \max\{U_2 , U_3 \}$. Then we say that the bivariate vector $(X_1 , X_2)$ has a bivariate
generalized exponential distribution with the shape parameters $\alpha_1$, $\alpha_2$ and $\alpha_3$ and the scale
parameter $\lambda$. We will denote it by BVGE($\alpha_1, \alpha_2, \alpha_3, \lambda$).  Now for the rest of the discussions
for brevity, we assume that $\lambda = 1$, although the results are true for general $\lambda$ also.  The
BVGE distribution with $\lambda = 1$ will be denoted by BVGE($\alpha_1, \alpha_2, \alpha_3$). Before providing the
joint CDF or PDF, we first mention how it may occur in practice.

 We know that if $U_{0,1}, U_{0,2}$ and $U_{0,3}$ are three independent random numbers, but follows GE($\alpha_{1}, \lambda_{1}$), GE($\alpha_{2}, \lambda_{1}$) and GE($\alpha_{3}, \lambda_{1}$) respectively, we can define $X_{1} = \max\{ U_{0,1}, U_{0,3}\}$,  $X_{2} = \max\{ U_{0,2}, U_{0,3}\}$ which follows BVGE($\alpha_{1}, \alpha_{2}, \alpha_{3}, \lambda_{1}$).

The joint cdf of BVGE can be written as :
\begin{eqnarray*}
F_{X_{1}, X_{2}}(x_{1}, x_{2})  & = &  F_{GE}(x_{1};\alpha_{1}, \lambda_{1})F_{GE}(x_{2};\alpha_{2}, \lambda_{1})F_{GE}(z;\alpha_{3}, \lambda_{1})\\
& = & F_{GE}(x_{1};\alpha_{1} + \alpha_{3}, \lambda_{1})F_{GE}(x_{2};\alpha_{2}, \lambda_{1}) ~~~~ \mbox{if} ~~ x_{1} < x_{2}\\
& = & F_{GE}(x_{1}; \alpha_{1}, \lambda_{1})F_{GE}(x_{2}; \alpha_{2} + \alpha_{3}, \lambda_{1}) ~~~~ \mbox{if} ~~ x_{1} > x_{2}\\
& = & F_{GE}(x; \alpha_{1} + \alpha_{2} + \alpha_{3}, \lambda_{1}) ~~~~ \mbox{if} ~~ x_{1} = x_{2} = x\\
\end{eqnarray*}
Therefore the joint pdf of $(X_{1}, X_{2})$ for $x_{1} > 0$ and $x_{2} > 0$, is :
\begin{eqnarray*}
f_{\alpha}(x_{1}, x_{2}) & = &  f_{1\alpha}(x_{1}, x_{2}) ~~ \mbox{if} ~~ 0 < x_{1} < x_{2} < \infty\\
& = & f_{2\alpha}(x_{1}, x_{2}) ~~ \mbox{if} ~~ 0 < x_{2} < x_{1} < \infty\\
& = & f_{0\alpha}(x) ~~ \mbox{if} ~~ 0 < x_{1} = x_{2} = x < \infty\\
\end{eqnarray*} where 
\begin{eqnarray*}
 &f_{1\alpha}&(x_{1}, x_{2}) =  f_{GE}(x_{1}; \alpha_{1} + \alpha_{3}, \lambda_{1}) f_{GE}(x_{2};\alpha_{2}, \lambda_{1})\\
 & = & (\alpha_{1} + \alpha_{3})\alpha_{2}(1 - e^{-\lambda_{1} x_{1}})^{\alpha_{1} + \alpha_{3} - 1}(1 - e^{-\lambda_{1} x_{2}})^{\alpha_{2} - 1} e^{-\lambda_{1} (x_{1} + x_{2})}\\
&f_{2\alpha}&(x_{1}, x_{2}) =   f_{GE}(x_{1}; \alpha_{1}, \lambda_{1}) f_{GE}(x_{2};\alpha_{2} + \alpha_{3}, \lambda_{1})\\
 & = & (\alpha_{1} + \alpha_{3})\alpha_{2}(1 - e^{-\lambda_{1} x_{1}})^{\alpha_{1} + \alpha_{3} - 1}(1 - e^{-\lambda_{1} x_{2}})^{\alpha_{2} - 1} e^{-\lambda_{1} (x_{1} + x_{2})}\\
&f_{0\alpha}&(x) =  \frac{\alpha_{3}}{\alpha_{1} + \alpha_{2} + \alpha_{3}}f_{GE}(x; \alpha_{1} + \alpha_{2} + \alpha_{3}, \lambda_{1}) 
\end{eqnarray*}

  Our aim is to study mixture of two bivariate generalized exponential distributions. Let BVGE($\alpha_{1}, \alpha_{2}, \alpha_{3}, \lambda_{1}$) and BVGE($\beta_{1}, \beta_{2}, \beta_{3}, \lambda_{2}$) be two independent bivariate generalized exponential distributions.  We consider mixture of them with mixture proportion $p_{0}$ and $p_{1}$.
\begin{eqnarray*}
f(x_{1}, x_{2}) = p_{0} f_{\alpha}(x_{1}, x_{2}; \alpha_{1}, \alpha_{2}, \alpha_{3}, \lambda_{1}) + p_{1} f_{\beta}(x_{1}, x_{2}; \beta_{1}, \beta_{2}, \beta_{3}, \lambda_{2})
\end{eqnarray*}

\begin{figure}[H]
 \begin{center}
  \subfigure[$\xi_{1}$]{\includegraphics[angle = -90, width = 0.45\textwidth]{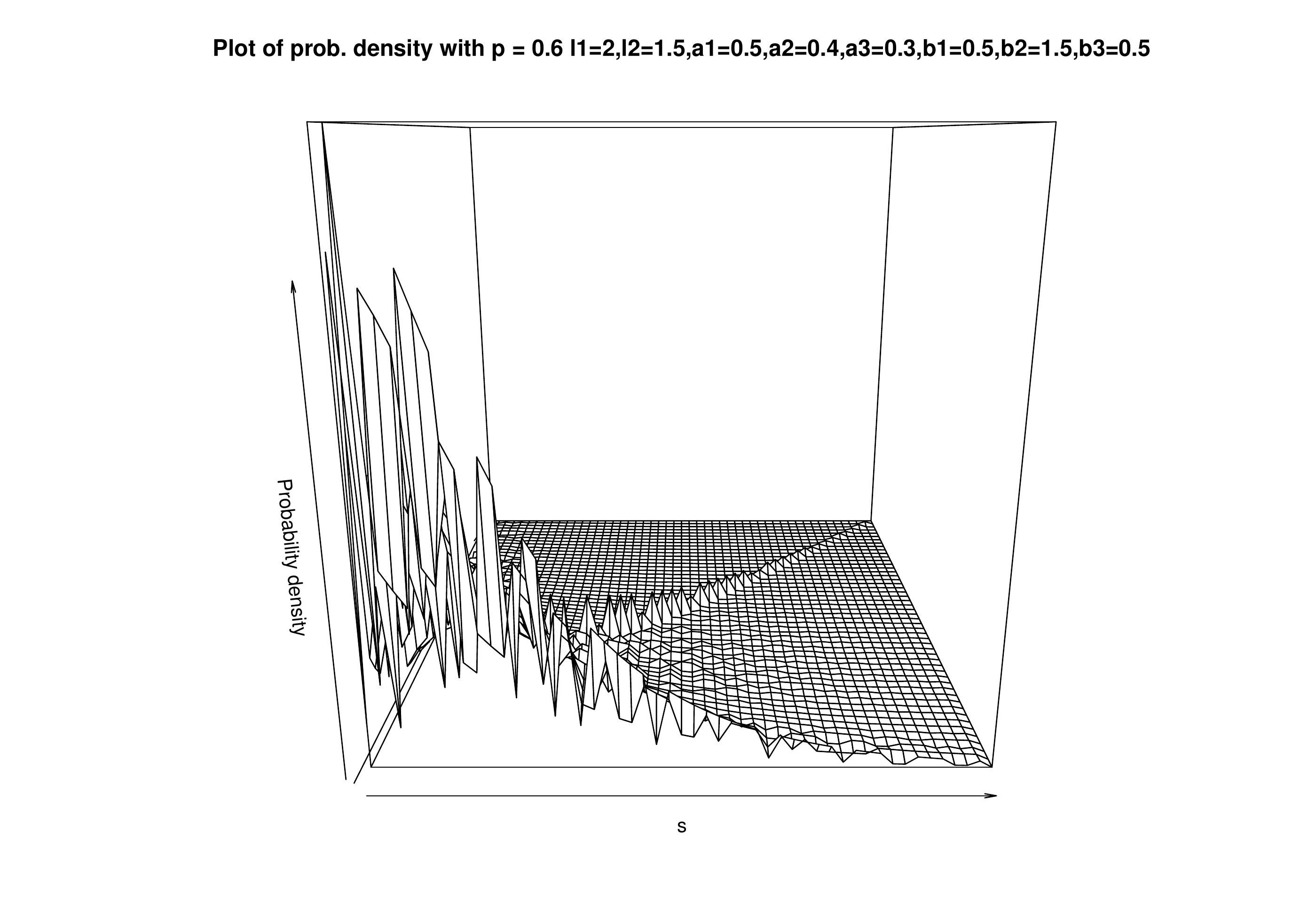}}
  \subfigure[$\xi_{2}$]{\includegraphics[angle = -90, width = 0.45\textwidth]{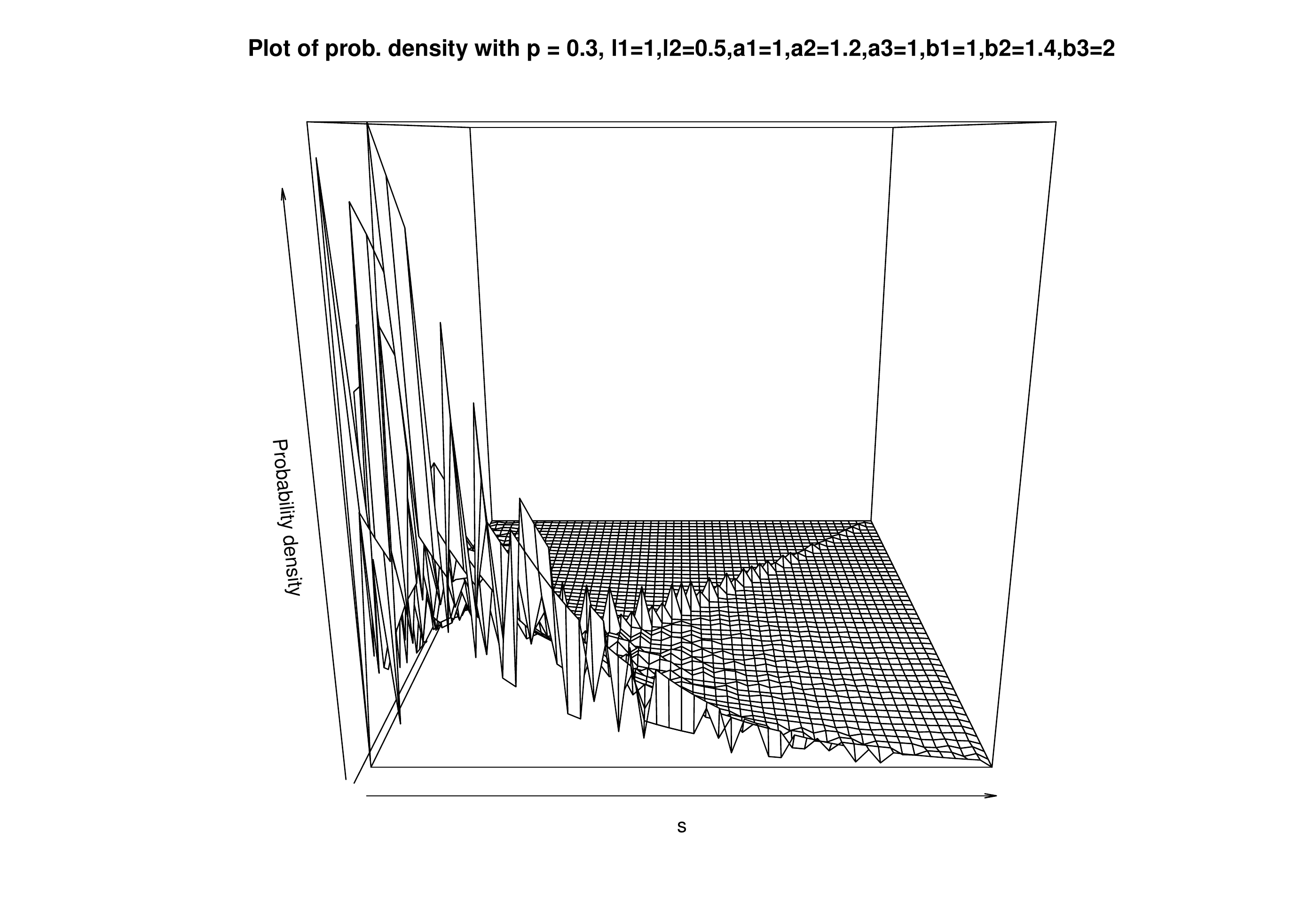}}\\
  \subfigure[$\xi_{1}$]{\includegraphics[angle = -90, width = 0.45\textwidth]{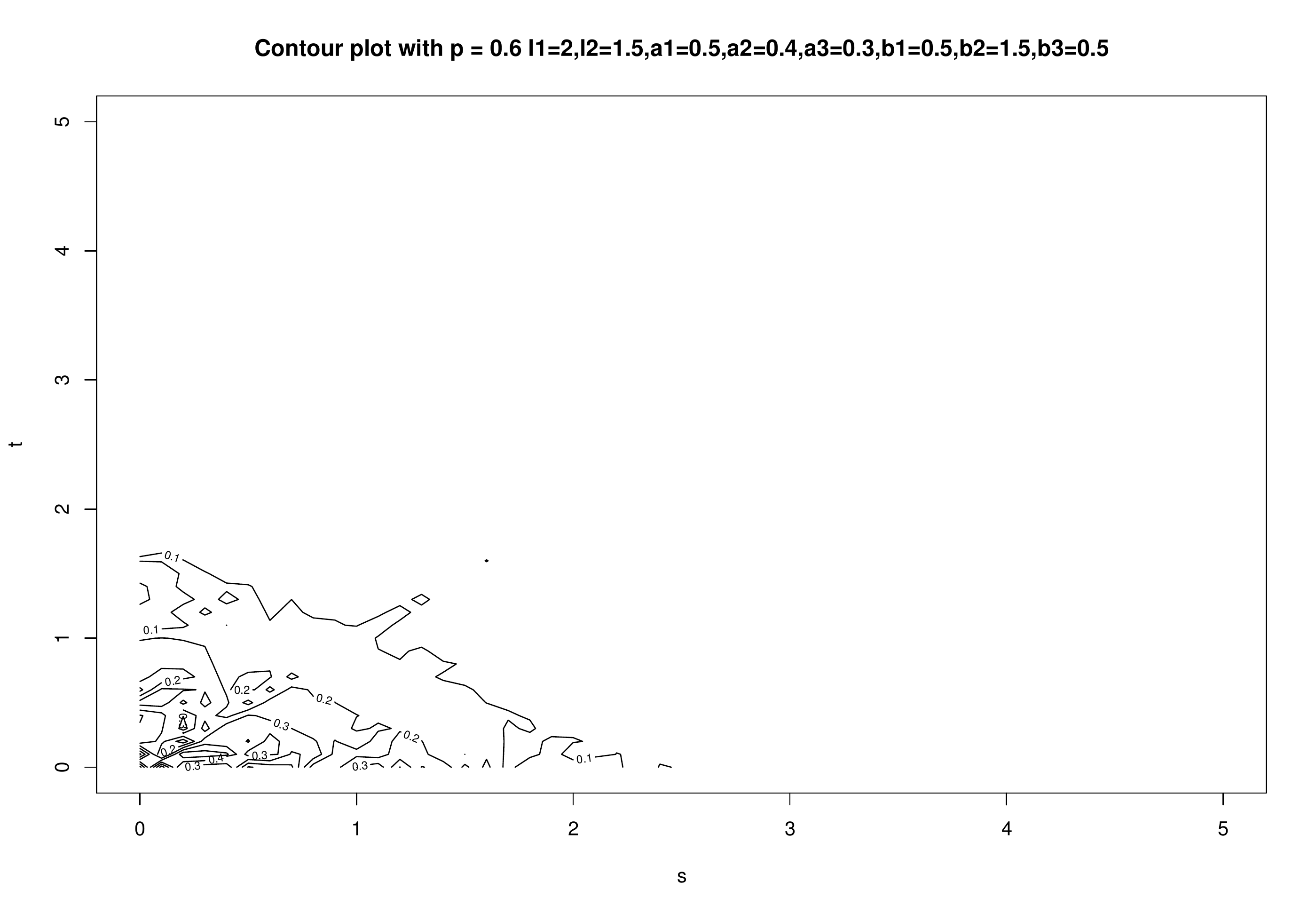}}
  \subfigure[$\xi_{2}$]{\includegraphics[angle = -90, width = 0.45\textwidth]{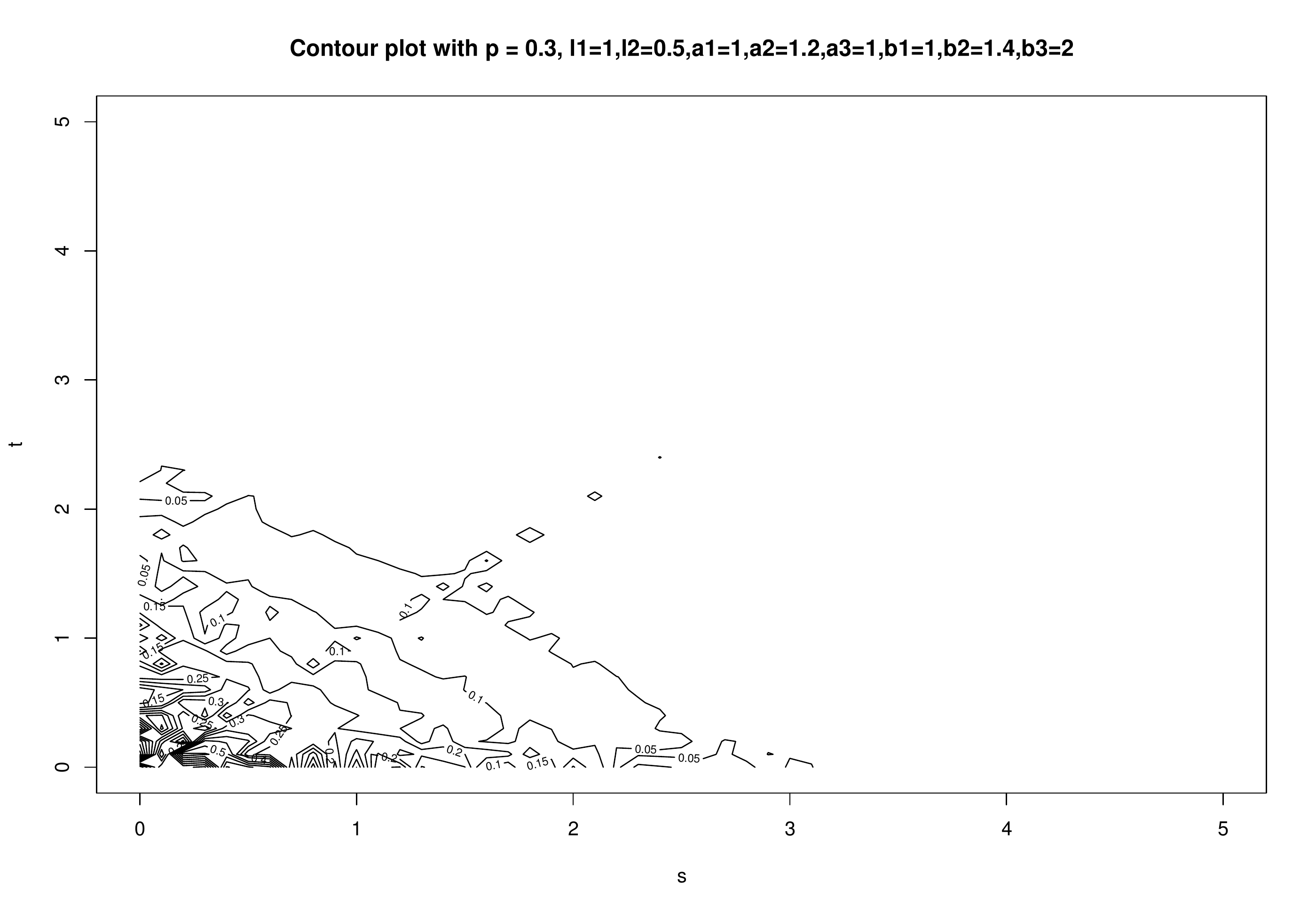}}\\
\caption{Surface and Contour plots of probability density function for different sets of parameters of MBVGE \label{fig1}}
\end{center}
\end{figure}

Figure-\ref{fig1} shows surface and contour plots of probability density function for four different sets of parameters of MBVGE.  They are as follows :
$\xi_{1}$ : $p = 0.6, \lambda_{1} = 2, \lambda_{2} = 1.5, \alpha_{1} = 0.5, \alpha_{2} = 0.4, \alpha_{3} = 0.3, \beta_{1} = 0.5, \beta_{2} = 1.5, \beta_{3} = 0.5$
$\xi_{2}$ : $p = 0.3, \lambda_{1} = 1, \lambda_{2} = 0.5, \alpha_{1} = 1, \alpha_{2} = 1.2, \alpha_{3} = 1, \beta_{1} = 1, \beta_{2} = 1.4, \beta_{3} = 2$.

\section{Properties}

\begin{theorem}
\begin{enumerate}
\item  Marginal distribution of MBVGE is mixture of univariate generalized exponential distribution.
\end{enumerate}
\end{theorem}
\begin{theorem} 
Copular function of MBVGE can be written as mixture of two different copulas i.e.:
$ C(u, v) = p C_{1}(u, v) + (1 - p) C_{2}(u, v) $ where $C_{1}(u, v)$ and $C_{2}(u, v)$ are two different copula and can be provided by the following expressions :
\begin{eqnarray*}
C_{1}(u, v) = \begin{cases} u^{\frac{\alpha_1}{\alpha_{1} + \alpha_{3}}} v & \text{$u^{\frac{1}{\alpha_{1} + \alpha_{3}}} <= v^{\frac{1}{\alpha_{2} + \alpha_{3}}}$} \\ u v^{\frac{\alpha_{2}}{\alpha_{2} + \alpha_{3}}} & \text{$u^{\frac{1}{\alpha_{1} + \alpha_{3}}} > v^{\frac{1}{\alpha_{2} + \alpha_{3}}}$} 
\end{cases}
\end{eqnarray*}
\begin{eqnarray*}
C_{2}(u, v) = \begin{cases} u^{\frac{\beta_{1}}{\beta_{1} + \beta_{3}}} v & \text{$u^{\frac{1}{\beta_{1} + \beta_{3}}} <= v^{\frac{1}{\beta_{2} + \beta_{3}}}$} \\ u v^{\frac{\beta_{2}}{\beta_{2} + \beta_{3}}} & \text{$u^{\frac{1}{\beta_{1} + \beta_{3}}} > v^{\frac{1}{\beta_{2} + \beta_{3}}}$} \end{cases}
\end{eqnarray*}
\end{theorem}
\begin{theorem} 
Tail Index of the Copula can be provided by the following relation :
\begin{eqnarray*}
\lambda_{L} & = & P[ Y < F^{-1}_{Y}(t) | X < F^{-1}_{X}(t)] = \lim_{t \rightarrow 0}\frac{C(t, t)}{t} = 0 \end{eqnarray*}
\begin{eqnarray*}
\lambda_{U} & = & P[ Y > F^{-1}_{Y}(t) | X > F^{-1}_{X}(t)]\\ & = & 2 - \lim_{t \rightarrow 1-} \frac{1 - C(t, t)}{1 - t}\\ & = &  2 - p \frac{\alpha_2}{\alpha_1 + \alpha_2 + \alpha_3} - (1 - p) \frac{\beta_2}{\beta_1 + \beta_2 + \beta_3}
\end{eqnarray*}
\end{theorem}
\begin{theorem} 
Hazard function for the distribution can be obtained from the relation : 
\begin{eqnarray*}
h(t_{1}, t_{2}) = \frac{f(t_{1}, t_{2})}{S(t_{1}, t_{2})} = \begin{cases} \frac{pf_{1\alpha}(x_{1}, x_{2}) + (1 - p)f_{1\beta}(x_{1}, x_{2})}{pS_{1\alpha}(x_{1}, x_{2}) + (1 - p)S_{1\beta}(x_{1}, x_{2})}  &  ~~~~~ \text{if ~~~~ $x_{1} < x_{2}$} \\
 \frac{pf_{2\alpha}(x_{1}, x_{2}) + (1 - p)f_{2\beta}(x_{1}, x_{2})}{pS_{2\alpha}(x_{1}, x_{2}) + (1 - p)S_{2\beta}(x_{1}, x_{2})}  &  ~~~~~ \text{if ~~~~ $x_{1} > x_{2}$} \\
 \frac{pf_{0\alpha}(x) + (1 - p)f_{0\beta}(x)}{pS_{0\alpha}(x, x) + (1 - p)S_{0\beta}(x, x)}  &  ~~~~~ \text{if ~~~~~ $x_{1} = x_{2} = x$} \end{cases}
\end{eqnarray*}
\end{theorem}
Other definition of Hazard function for the distribution can be obtained from the relation : 
\begin{eqnarray*}
h(t_{1}, t_{2}) =  [\frac{-\partial}{\partial t_{1}} \ln S(t_{1}, t_{2}), \frac{-\partial}{\partial t_{2}} \ln S(t_{1}, t_{2})]
\end{eqnarray*}

\begin{theorem}
\begin{eqnarray*}
P(X_{1} \leq x_{1} | X_{2} = x_{2}) = \begin{cases} \frac{pA_{1} + (1 - p)B_{1}}{pA + (1 - p)B} & ~~\text{$x_{1} < x_{2}$}\\
 \frac{pC_{1} + (1 - p)D_{1}}{pA + (1 - p)B} & ~~\text{$x_{1} > x_{2}$}\\
 \frac{pE_{1} + (1 - p)F_{1}}{pA + (1 - p)B} & ~~ \text{$x_{1} = x_{2}$} \end{cases}
\end{eqnarray*}
where $ A_{1} = (1 - e^{-\lambda x_{1}})^{\alpha_1 + \alpha_3}\alpha_2(1 - e^{-\lambda x_{2}})^{\alpha_2 -1}$, $B_{1} = (1 - e^{-\lambda x_{1}})^{\beta_1 + \beta_3}\beta_2(1 - e^{-\lambda x_{2}})^{\beta_2 -1}$, $C_{1} = (1 - e^{-\lambda x_{1}})^{\alpha_1}(\alpha_2 + \alpha_3)(1 - e^{-\lambda x_{2}})^{\alpha_2 + \alpha_3 - 1}$, $D_{1} = (1 - e^{-\lambda x_{1}})^{\beta_1}(\beta_2 + \beta_3)(1 - e^{-\lambda x_{2}})^{\beta_2 + \beta_3 - 1}$, $E_{1} = (\alpha_1 + \alpha_2 + \alpha_3)(1 - e^{-\lambda x_{2}})^{\alpha_1 + \alpha_2 + \alpha_3 - 1}$, $F_{1} = (\beta_{1} + \beta_{2} + \beta_{3})(1 - e^{-\lambda x_{2}})^{\beta_1 + \beta_2 + \beta_3 -1}$, $ A = (\alpha_2 + \alpha_3)(1 - e^{-\lambda x_{2}})^{\alpha_2 + \alpha_3 - 1}$, $B = (\beta_2 + \beta_3)(1 - e^{-\lambda x_{2}})^{\beta_2 + \beta_3 - 1}$. 
\end{theorem}

\begin{theorem}
Expression for Kendal's tau can be obtained using its copula form as
\begin{eqnarray*}
\tau & = &  p^{2}\frac{(\alpha_1 + \alpha_2)}{(\alpha_1 + \alpha_2 + \alpha_3)} + (1 - p)^{2}\frac{\beta_1 + \beta_2}{(\beta_1 + \beta_2 + \beta_3)}\\ & + & 2p(1 - p)\frac{\beta_2 (\alpha_2 + \alpha_3)}{(2\beta_1 + \beta_2 + 2\beta_3)(\alpha_ 2 + \alpha_3) + \alpha_2(\beta_2 + \beta_3)} \\ & + & 2p(1 - p)\frac{\alpha_2 (\beta_2 + \beta_3)}{(2\alpha_1 + \alpha_2 + 2\alpha_3)(\beta_2 + \beta_3) + \beta_2(\alpha_2 + \alpha_3)}\\ & + & 2p(1 - p)\frac{\beta_1(\alpha_1 + \alpha_3)}{2(\beta_1 + \beta_2 + \beta_3)\alpha_1 + (2\beta_2 + \beta_1 + \beta_3)\alpha_3} \\ & + & 2p(1 - p)\frac{\alpha_1(\beta_1 + \beta_3)}{2(\alpha_1 + \alpha_2 + \alpha_3)\beta_1 + (2\alpha_2 + \alpha_1 + \alpha_3)\beta_3} - 1
\end{eqnarray*}
\end{theorem}
Similarly,
\begin{theorem}
Expression for Spearman Correlation coefficient ($r_{s}$) can be provided as
\begin{eqnarray*}
r_{s} = 6p\frac{\alpha_1 + \alpha_2}{2(\alpha_1 + \alpha_2 + \alpha_3) + \alpha_3} + 6(1 - p)\frac{(\beta_ 1 + \beta_2)}{2(\beta_ 1 + \beta_2 + \beta_3) + \beta_3} - 3.
\end{eqnarray*}
\end{theorem}

\section{Implementation of EM algorithm}

  Here we use multistage EM algorithm to construct the final pseudo-likelihood.  In stage -1, we introduce 
\begin{eqnarray*} z_{i} =
\begin{cases} 1 & ~~~ \text{if ~~~ $(x_{1i}, x_{2i}) \sim f_{\alpha}(x_{1i}, x_{2i})$}\\ 0 & ~~~ \text{if ~~~ $(x_{1i}, x_{2i}) \sim f_{\beta}(x_{1i}, x_{2i})$}\end{cases}
\end{eqnarray*}

  Depending on observations lying on $I_{0}, I_{1}$ and $I_{2}$, we can define three parts of posterior distribution of $Z_{i}$, as $p_{0, 0i}$, $p_{0, 1i}$ and $p_{0, 2i}$ respectively.

  Therefore, \begin{eqnarray*} p_{0, 0i} & = & P(Z_{i} = 1 | (X_{1}, X_{2}) \in I_{0})\\ & = & \frac{p_{0}f_{0\alpha}(x_{1}, x_{2})}{p_{0}f_{0\alpha}(x_{1}, x_{2}) + p_{1}f_{0\beta}(x_{1}, x_{2})} ~~~ \mbox{for $i = 1, \cdots, n_{0}$} \end{eqnarray*}
\begin{eqnarray*} p_{0, 1i} & = & P(Z_{i} = 1 | (X_{1}, X_{2}) \in I_{1})\\ & = & \frac{p_{0}f_{1\alpha}(x_{1}, x_{2})}{p_{0}f_{1\alpha}(x_{1}, x_{2}) + p_{1}f_{1\beta}(x_{1}, x_{2})} ~~~ \mbox{for}~~ i = 1, \cdots, n_{1}  \end{eqnarray*}
\begin{eqnarray*} p_{0, 2i} & = & P(Z_{i} = 1 | (X_{1}, X_{2}) \in I_{2})\\ & = & \frac{p_{0}f_{2\alpha}(x_{1}, x_{2})}{p_{0}f_{2\alpha}(x_{1}, x_{2}) + p_{1}f_{2\beta}(x_{1}, x_{2})} ~~~ \mbox{for} i = 1, \cdots, n_{2}  \end{eqnarray*}

  We also take $p_{1,0i} = (1 - p_{0,0i}), p_{1,1i} = (1 - p_{0,1i})$ and $p_{1, 2i} = (1 - p_{0,2i})$.

  In second stage we take the missing information as the maximum between the the pair of observations corresponding to $(X_{1}, X_{2})$.  Therefore we introduce $(\Delta_{0,0}, \Delta_{0,1})$ if we assume $(X_{1}, X_{2}) \sim f_{\alpha}(\cdot)$ and $(\Delta_{1,0}, \Delta_{1,1})$ for $(X_{1}, X_{2}) \sim f_{\alpha}(\cdot)$ as described in \cite{GuptaKundu:2009} i.e. $\Delta_{1,0} = 1 ~~ \mbox{or} ~~ 3$  if $U_{0,1} > U_{0,3}$ or $U_{0,1} < U_{0,3}$ and $\Delta_{0,1} = 2 ~~\mbox{or}~~ 3$ if $U_{0,2} > U_{0,3}$ or $U_{0,2} < U_{0,3}$.  If $\gamma_{1} = (\alpha_{1}, \alpha_{2}, \alpha_{3}, \lambda_{1})$, fractional mass $(u_{0,1}(\gamma), u_{0,2}(\gamma))$ [We denote simply as $u_{0,1}$, $u_{0,2}$] assign to 'pseudo observation' $(x_{1}, x_{2})$ is the conditional probability that the random vector $(\Delta_{0,1}, \Delta_{0,2})$ takes the values $(1,2)$ or $(3,2)$ respectively given that $X_{1} < X_{2}$.

  Similarly, if $(x_{1}, x_{2}) \in I_{2}$, we form the pseudo observations by introducing fractional mass $w_{0,1}$ and $w_{0,2}$ which is the conditional distribution that the random vector $(\Delta_{0,0}, \Delta_{0,1})$ takes the values (1, 2) and (1, 3) respectively, given that $X_{1} > X_{2}$.

  We can show $u_{0,1} = \frac{\alpha_{1}}{\alpha_{1} + \alpha_{3}}$ and $u_{0,2} = \frac{\alpha_{3}}{\alpha_{1} + \alpha_{3}}$ whereas $w_{0,1} = \frac{\alpha_{2}}{\alpha_{2} + \alpha_{3}}$ and $w_{0,2} = \frac{\alpha_{3}}{\alpha_{2} + \alpha_{3}} $

  Exactly in the similar line we can define $(\Delta_{1, 0}, \Delta_{1, 1})$ for second type of bivariate generalized exponential distribution and we denote four conditional probabilities as ($u_{1,1}$, $u_{1,2}$) and ($w_{1,1}$, $w_{1,2}$) where $u_{1,1} = \frac{\beta_{1}}{\beta_{1} + \beta_{3}}$, $u_{1,2} = \frac{\beta_{3}}{\beta_{1} + \beta_{3}}$, $w_{1,1} = \frac{\beta_{2}}{\beta_{2} + \beta_{3}}$ and $w_{1,2} = \frac{\beta_{3}}{\beta_{2} + \beta_{3}}$. 

Therefore first step log-likelihood can be written as 
\begin{eqnarray*}
\mathcal{L}  =  \log \prod_{i = 1}^{n} [p f_{\alpha}(x_{1i},x_{2i})]^{z_{i}} [(1 - p)f_{\beta}(x_{1i}, x_{2i})]^{1 - z_{i}}
\end{eqnarray*}

  In the second step we use complete information in $\log(f_{\alpha}(x_{1i}, x_{2i}))$ and $\log(f_{\beta}(x_{1i}, x_{2i}))$  by introducing $(\Delta_{0,0}, \Delta_{0,1})$  and $(\Delta_{1,0}, \Delta_{1,1})$ respectively.  In the calculation of pseudo-likelihood we only need to take care of the proper usage of posterior of $z_{i}$ given the data $(x_{1i}, x_{2i})$.  Formulation of the E-step and M-step is shown in the subsequent subsections.

\subsection{Formulation of E-step}

The form of pseudo-likelihood can be written as follows :
\begin{eqnarray*} && l_{pseudo}(\alpha_{1}, \alpha_{2}, \alpha_{3}, \lambda_{1}, \beta_{1}, \beta_{2}, \beta_{3}, \lambda_{2})\\ & = & \mbox{contribution from 1st part of pseudo likelihood}\\ & + & \mbox{contribution from 2nd part of pseudo likelihood} \end{eqnarray*}
\begin{eqnarray*}
&& \mbox{First part of pseudo likelihood}\\ & = &  [\sum_{i = 1}^{n_{0}}p_{0,0i} + \sum_{i = 1}^{n_{1}}p_{0,1i} + \sum_{i = 1}^{n_{2}}p_{0,2i})]\ln p + [\sum_{i = 1}^{n_{0}}p_{1,0i} + \sum_{i = 1}^{n_{1}}p_{1,1i} + \sum_{i = 1}^{n_{2}}p_{1,2i})]\ln (1 - p)\\ & + & \sum_{i \in I_{0}}^{} p_{0,0i}\ln(\alpha_{3}) + \sum_{i \in I_{0}}^{} p_{0,0i} \ln \lambda_{1} + (\alpha_{1} + \alpha_{2} + \alpha_{3} - 1)\sum_{i \in I_{0}}^{} p_{0,0i} \ln(1 - e^{-\lambda_{1} y_{i}})\\ & - & \lambda_{1} \sum_{i \in I_{0}}^{} p_{0,0i} x_{1i} + u_{0,1}[\sum_{i \in I_{1}}^{} p_{0,1i} \ln\alpha_{1} + 2 \sum_{i \in I_{1}}^{}p_{0,1i} \ln \lambda_{1} - \lambda_{1} \sum_{i \in I_{1}}^{} p_{0,1i} x_{1i}\\ & + & (\alpha_{1} + \alpha_{3} - 1)\sum_{i \in I_{1}}^{} p_{0,1i}\ln (1 - e^{-\lambda_{1} x_{1i}}) ]\\ & + & u_{0,2}[ \sum_{i \in I_{1}}^{} p_{0,1i} \ln\alpha_{3} + 2\sum_{i \in I_{1}}^{} p_{0,1i}\ln\lambda_{1} - \lambda_{1} \sum_{i \in I_{1}}^{} p_{0,1i} x_{1i}\\ & + & (\alpha_{1} + \alpha_{3} - 1)\sum_{i \in I_{1}}^{} p_{0,1i}\ln(1 - e^{-\lambda_{1} x_{1i}})] + [\sum_{i \in I_{1}}^{} p_{0,1i}\ln\alpha_{2}\\ & - & \lambda_{1}\sum_{i \in I_{1}}^{}p_{0,1i}x_{2i} + (\alpha_{2} - 1)\sum_{i \in I_{1}}^{}p_{0,1i}\ln(1 - e^{-\lambda_{1} x_{2i}})] + w_{0,1}[\sum_{i \in I_{2}}^{} p_{0,2i}\ln\alpha_{2} \\ & + & 2\sum_{i \in I_{2}}^{} p_{0,2i}\ln\lambda_{1} - \lambda_{1} \sum_{i \in I_{2}}^{} p_{0,2i}x_{2i} + (\alpha_{2} + \alpha_{3} - 1)\sum_{i \in I_{2}}^{} p_{0,2i}\ln(1 - e^{-\lambda_{1} x_{2i}})]\\ & + & w_{0,2}[\sum_{i \in I_{2}}^{}p_{0,2i}\ln\alpha_{3} + 2\sum_{i \in I_{2}}^{} p_{0,2i}\ln\lambda_{1} - \lambda_{1} \sum_{i \in I_{2}}^{} p_{0,2i}x_{2i}\\ & + &  (\alpha_{2} + \alpha_{3} - 1)\sum_{i \in I_{2}}^{} p_{0,2i}\ln(1 - e^{-\lambda_{1} x_{2i}})] + [\sum_{i \in I_{2}}^{} p_{0,2i}\ln\alpha_{1}\\ & - & \lambda_{1} \sum_{i \in I_{2}}^{} p_{0,2i}x_{1i} + (\alpha_{1} - 1)\sum_{i \in I_{2}}^{} p_{0,2i}\ln(1 - e^{-\lambda_{1} x_{1i}})]
\end{eqnarray*}

\begin{eqnarray*}
&& \mbox{Second part of pseudo likelihood}\\ & + & \sum_{i \in I_{0}}^{} p_{1,0i}\ln(\beta_{3}) + \sum_{i \in I_{0}}^{} p_{1,0i} \ln \lambda_{2} + (\beta_{1} + \beta_{2} + \beta_{3} - 1)\sum_{i \in I_{0}}^{} p_{1,0i} \ln(1 - e^{-\lambda_{2} y_{i}})\\ & - & \lambda_{2} \sum_{i \in I_{0}}^{} p_{1,0i} x_{1i} + u_{1,1}[\sum_{i \in I_{1}}^{} p_{1,1i} \ln\beta_{1} + 2 \sum_{i \in I_{1}}^{}p_{1,1i} \ln \lambda_{2} - \lambda_{2} \sum_{i \in I_{1}}^{} p_{1,1i} x_{1i}\\ & + & (\beta_{1} + \beta_{3} - 1)\sum_{i \in I_{1}}^{} p_{1,1i}\ln (1 - e^{-\lambda_{2} x_{1i}}) ]\\ & + & u_{1,2}[ \sum_{i \in I_{1}}^{} p_{1,1i} \ln\beta_{3} + 2\sum_{i \in I_{1}}^{} p_{1,1i}\ln\lambda_{2} - \lambda_{2} \sum_{i \in I_{1}}^{} p_{1,1i} x_{1i}\\ & + & (\beta_{1} + \beta_{3} - 1)\sum_{i \in I_{1}}^{} p_{1,1i}\ln(1 - e^{-\lambda_{2} x_{1i}})] + [\sum_{i \in I_{1}}^{} p_{1,1i}\ln\beta_{2}\\ & - & \lambda_{2}\sum_{i \in I_{1}}^{}p_{1,1i}x_{2i} + (\beta_{2} - 1)\sum_{i \in I_{1}}^{}p_{1,1i}\ln(1 - e^{-\lambda_{2} x_{2i}})] + w_{1,1}[\sum_{i \in I_{2}}^{} p_{1,2i}\ln\beta_{2} \\ & + & 2\sum_{i \in I_{2}}^{} p_{1,2i}\ln\lambda_{2} - \lambda_{2} \sum_{i \in I_{2}}^{} p_{1,2i}x_{2i} + (\beta_{2} + \beta_{3} - 1)\sum_{i \in I_{2}}^{} p_{1,2i}\ln(1 - e^{-\lambda_{2} x_{2i}})]\\ & + & w_{1,2}[\sum_{i \in I_{2}}^{} p_{1,2i}\ln\beta_{3} + 2\sum_{i \in I_{2}}^{} p_{1,2i}\ln\lambda_{2} - \lambda_{2} \sum_{i \in I_{2}}^{} p_{1,2i}x_{2i}\\ & + & (\beta_{2} + \beta_{3} - 1)\sum_{i \in I_{2}}^{} p_{1,2i}\ln(1 - e^{-\lambda_{2} x_{2i}})] + [\sum_{i \in I_{2}}^{} p_{1,2i}\ln\beta_{1}\\ & - & \lambda_{2} \sum_{i \in I_{2}}^{} p_{1,2i}x_{1i} + (\beta_{1} - 1)\sum_{i \in I_{2}}^{} p_{1,2i}\ln(1 - e^{-\lambda_{2} x_{1i}}) ] 
\end{eqnarray*}

\subsection{Formulation of M-step:}

  Now the `M' step involves the maximization of the $$l_{pseudo}(\alpha_{1}, \alpha_{2}, \alpha_{3}, \lambda_{1}, \beta_{1}, \beta_{2}, \beta_{3}, \lambda_{2}, p)$$ with respect to all parameters.  Taking derivative with respect to $p$, yields, 
$$ p = \frac{\sum_{i =1 }^{n_{0}} p_{0,0i} + \sum_{i = 1}^{n_{1}} p_{0,1i} + \sum_{i = 1}^{n_{2}} p_{0,2i}}{n} $$

  For fixed $\lambda_{1}$ and $\lambda_{2}$, the maximization of $l_{pseudo}(\cdot)$ occurs at 
$$\small\small \hat{\alpha}_{1}(\lambda_{1}) = -\frac{u_{0,1} \sum_{i \in I_{1}}^{} p_{0,1i} + \sum_{i \in I_{2}}^{} p_{0,2i}}{A} $$

$ A = \sum_{i \in I_{0}}^{} p_{0,0i}\ln(1 - e^{-\lambda_{1} x_{i}}) +  \sum_{i \in I_{1}}^{} p_{0,1i} \ln(1 - e^{-\lambda_{1} x_{1i}}) + \sum_{i \in I_{2}}^{}p_{0,2i}\ln(1 - e^{-\lambda_{1} x_{1i}}) $
$$\small\small \hat{\alpha}_{2}(\lambda_{1}) = \frac{-\sum_{i \in I_{1}}^{} p_{0,1i} - w_{0,1}\sum_{i \in I_{2}}^{} p_{0,2i}}{B} $$  
$B = \sum_{i \in I_{0}}^{} p_{0,0i}\ln(1 - e^{-\lambda_{1} x_{i}}) + \sum_{i \in I_{1}}^{} p_{0,1i}\ln(1 - e^{-\lambda_{1} x_{2i}}) + \sum_{i \in I_{2}}^{} p_{0,2i} \ln(1 - e^{-\lambda_{1} x_{2i}}) $  
$$\small\small \hat{\alpha}_{3}(\lambda_{1}) = \frac{-\sum_{i \in I_{0}}^{} p_{0,0i} - u_{0,2}\sum_{i \in I_{1}}^{} p_{0,1i} - w_{0,2}\sum_{i \in I_{2}}^{} p_{0,2i} }{D} $$
$ D = \sum_{i \in I_{0}}^{} p_{0,0i}\ln(1 - e^{-\lambda_{1} x_{i}}) + \sum_{i \in I_{1}}^{} p_{0,1i}\ln(1 - e^{-\lambda_{1} x_{1i}}) + \sum_{i \in I_{2}}^{} p_{0,2i} \ln(1 - e^{-\lambda_{1} x_{2i}}) $
and $\hat{\lambda}_{1}$, which maximizes $l_{pesudo}(\cdot)$ can be obtained as a solution of the following fixed point equation;
$$ g_{1}(\lambda_{1}) =\lambda_{1} $$
 
where \begin{eqnarray*} g_{1}(\lambda_{1}) & = &  \frac{E_{1}}{F_{1}}   \end{eqnarray*}
\begin{eqnarray*} E_{1} = \sum_{i \in I_{0}}^{} p_{0,0i}  + 2 u_{0,1} \sum_{i \in I_{1}}^{} p_{0,1i} + 2 u_{0,2} \sum_{i \in I_{1}}^{} p_{0,1i} + 2w_{0,1}\sum_{i \in I_{2}}^{} p_{0,2i} + 2w_{0,2}\sum_{i \in I_{2}}^{} p_{0,2i} \end{eqnarray*}
\begin{eqnarray*} F_{1} & = & \sum_{i \in I_{0}}^{} p_{0,0i} x_{1i} - (\alpha_{1} + \alpha_{2} + \alpha_{3} - 1)\sum_{i \in I_{0}}^{} \frac{y_{i}p_{0,0i}e^{-\lambda_{1}y_{i}}}{(1 - e^{-\lambda_{1}y_{i}})}\\ & + & u_{0,1}\sum_{i \in I_{1}}^{} p_{0,1i}x_{1i} - u_{0,1}(\alpha_{1} + \alpha_{3} - 1)\sum_{i \in I_{1}}^{}p_{0,1i} x_{1i}\frac{e^{-\lambda_{1} x_{1i}}}{(1 - e^{-\lambda_{1}x_{1i}})}\\ & + &  u_{0,2} \sum_{i \in I_{1}}^{} p_{0,1i}x_{1i} - u_{0,2} (\alpha_{1} + \alpha_{3} - 1)\sum_{i \in I_{1}}^{}p_{0,1i} \frac{x_{1i}e^{-\lambda_{1} x_{1i}}}{(1 - e^{-\lambda_{1} x_{1i}})} \\ & + & w_{0,1}\sum_{i \in I_{2}}^{}p_{0,2i} x_{2i} - w_{0,1}(\alpha_{2} + \alpha_{3} - 1)\sum_{i \in I_{2}}^{}p_{0,2i}\frac{x_{2i}e^{-\lambda_{1}x_{2i}}}{(1 - e^{-\lambda_{1}x_{2i}})}\\ & + & w_{0,2}\sum_{i \in I_{2}}^{} p_{0,2i} x_{2i} - w_{0,2}(\alpha_{2} + \alpha_{3} - 1)\sum_{i \in I_{2}}^{} p_{0,2i}\frac{x_{2i}e^{-\lambda_{1} x_{2i}}}{(1 - e^{-\lambda_{1} x_{2i}})}\\ & + & \sum_{i \in I_{2}}^{} p_{0,2i}x_{1i} - (\alpha_{1} - 1)\sum_{i \in I_{2}}^{} \frac{p_{0,2i}x_{1i}e^{-\lambda_{1} x_{1i}}}{(1 - e^{-\lambda_{1} x_{1i}})}\\ & + & \sum_{i \in I_{1}}^{} p_{0,1i}x_{2i} - (\alpha_{2} - 1)\sum_{i \in I_{1}}^{}p_{0,1i}\frac{x_{2i}e^{-\lambda_{1} x_{2i}}}{(1 - e^{-\lambda_{1} x_{2i}})} \end{eqnarray*}
 
Similarly from the second part, we get 

$$\small\small \hat{\beta}_{1}(\lambda_{2}) = -\frac{u_{1,1} \sum_{i \in I_{1}}^{} p_{1,1i} + \sum_{i \in I_{2}}^{} p_{1,2i}}{G} $$
$  G = \sum_{i \in I_{0}}^{} p_{1,0i}\ln(1 - e^{-\lambda_{2} x_{i}}) + \sum_{i \in I_{1}}^{} p_{1,1i} \ln(1 - e^{-\lambda_{2} x_{1i}}) + \sum_{i \in I_{2}}^{}p_{1,2i}\ln(1 - e^{-\lambda_{2} x_{1i}}) $
$$\small\small \hat{\beta}_{2}(\lambda_{2}) = \frac{-\sum_{i \in I_{1}}^{} p_{1,1i} - w_{1,1}\sum_{i \in I_{2}}^{} p_{1,2i}}{H} $$  
$ H = \sum_{i \in I_{0}}^{} p_{1,0i}\ln(1 - e^{-\lambda_{2} x_{i}}) + \sum_{i \in I_{1}}^{} p_{1,1i}\ln(1 - e^{-\lambda_{2} x_{2i}}) + \sum_{i \in I_{2}}^{} p_{1,2i} \ln(1 - e^{-\lambda_{2} x_{2i}}) $
$$\small\small \hat{\beta}_{3}(\lambda_{2}) = \frac{-\sum_{i \in I_{0}}^{} p_{1,0i} - u_{1,2}\sum_{i \in I_{1}}^{} p_{1,1i} - w_{1,2}\sum_{i \in I_{2}}^{} p_{1,2i} }{K} $$
$ K = \sum_{i \in I_{0}}^{} p_{1,0i}\ln(1 - e^{-\lambda_{2} x_{i}}) + \sum_{i \in I_{1}}^{} p_{1,1i}\ln(1 - e^{-\lambda_{2} x_{1i}}) + \sum_{i \in I_{2}}^{} p_{1,2i} \ln(1 - e^{-\lambda_{2} x_{2i}}) $

$\hat{\lambda}_{2}$, which maximizes $l_{pesudo}(\cdot)$ can be obtained as a solution of the following fixed point equation;
$$ g_{1}(\lambda_{2}) = \lambda_{2} $$
 
where \begin{eqnarray*} g_{1}(\lambda_{2}) & = &  \frac{K_{1}}{L_{1}}   \end{eqnarray*}

$$K_{1} = \sum_{i \in I_{0}}^{} p_{1,0i}  + 2 u_{1,1} \sum_{i \in I_{1}}^{} p_{1,1i} + 2 u_{1,2} \sum_{i \in I_{1}}^{} p_{1,1i} + 2w_{1,1}\sum_{i \in I_{2}}^{} p_{1,2i} + 2w_{1,2}\sum_{i \in I_{2}}^{} p_{1,2i} $$

\begin{eqnarray*} L_{1} & = & \sum_{i \in I_{0}}^{} p_{1,0i} x_{1i} - (\beta_{1} + \beta_{2} + \beta_{3} - 1)\sum_{i \in I_{0}}^{} \frac{y_{i}p_{1,0i}e^{-\lambda_{2}y_{i}}}{(1 - e^{-\lambda_{2}y_{i}})}\\ & + & u_{1,1}\sum_{i \in I_{1}}^{} p_{1,1i}x_{1i} - u_{1,1}(\beta_{1} + \beta_{3} - 1)\sum_{i \in I_{1}}^{}p_{1,1i} x_{1i}\frac{e^{-\lambda_{2} x_{1i}}}{(1 - e^{-\lambda_{2}x_{1i}})}\\ & + &  u_{1,2} \sum_{i \in I_{1}}^{} p_{1,1i}x_{1i} - u_{1,2} (\beta_{1} + \beta_{3} - 1)\sum_{i \in I_{1}}^{}p_{1,1i} \frac{x_{1i}e^{-\lambda_{2} x_{1i}}}{(1 - e^{-\lambda_{2} x_{1i}})} \\ & + & \sum_{i \in I_{1}}^{} p_{1,1i}x_{2i} - (\beta_{2} - 1)\sum_{i \in I_{1}}^{}\frac{p_{1,1i}x_{2i}e^{-\lambda_{2} x_{2i}}}{(1 - e^{-\lambda_{2} x_{2i}})}\\  & + & w_{1,1}\sum_{i \in I_{2}}^{}p_{1,2i} x_{2i} - w_{1,1}(\beta_{2} + \beta_{3} - 1)\sum_{i \in I_{2}}^{}p_{1,2i}\frac{x_{2i}e^{-\lambda_{2}x_{2i}}}{(1 - e^{-\lambda_{2}x_{2i}})} + w_{1,2}\sum_{i \in I_{2}}^{} p_{1,2i} x_{2i} \\ & - & w_{1,2}(\beta_{2} + \beta_{3} - 1)\sum_{i \in I_{2}}^{} p_{1,2i}\frac{x_{2i}e^{-\lambda_{2} x_{2i}}}{(1 - e^{-\lambda_{2} x_{2i}})}\\ & + & \sum_{i \in I_{2}}^{} p_{1,2i}x_{1i} - (\beta_{1} - 1)\sum_{i \in I_{2}}^{} \frac{p_{1,2i}x_{1i}e^{-\lambda_{2} x_{1i}}}{(1 - e^{-\lambda_{2} x_{1i}})} \end{eqnarray*}

\section{Numerical Result}

  We use package R 3.2.3 to perform the estimation procedure.  All the programs will be available on request to author. We
have taken two different sets of parameters to conduct our simulation.
These are $\alpha_1 = 1; \alpha_2 = 1.2; \alpha_3 = 1 ; \beta_1 = 1; \beta_2 = 1.4; \beta_3 = 2 ; \lambda_1 = 1; \lambda_2 = 0.5; p = 0.3$.
$\alpha_1 = 0.5; \alpha_2 = 0.4; \alpha_3 = 0.3 ; \beta_1 = 0.5 ; \beta_2 = 1.5 ; \beta_3 = 0.5 ; \lambda_1 = 1; \lambda_2 = 0.5; p = 0.6$.

  We take sample size as $n = 1000, 1500$.  The procedure demands high sample size as it uses many stages of approximation.  We start EM algorithm with random initial guesses at each iteration.  Left side of the Table-\ref{AEMSE} shows the values of the parameters of parent distributions from which data is generated.  We use stopping criteria as absolute value of likelihood changes with respect to previous likelihood at each iteration.  The average estimates (AE), mean squared error (MSE) are reported based on 1000 replications.  With a very small probability, algorithm is unable to find out the convergent point under this stopping criteria.  As a remedy we stop the algorithm after 5000 iterations.  This won't affect the estimates and MSE much, because due to some reason, algorithm was unable to reach the convergence point.  However it will roam around the actual values. Since major objective of EM is to extract some closer value of the original parameters, we observe in our simulation experiment that the goal will achieve without much affecting average estimates and mean square error. In practice we can use other optimization techniques taking the initial values of the parameters as the values that we have obtained using EM algorithm to get more perfect estimates.    

\begin{table}[!h]
{\begin{tabular}[l]{@{}lccccc}\hline
Parameter Set &  & $\alpha_1 = 1$ & $\alpha_2 = 1.2$ & $\alpha_3 = 1$ \\ 
 n = 1500 & Average Estimates & 1.0902 & 1.2372 & 1.003 \\ 
 & Mean Square Error & 0.1047 & 0.0223 & 0.0119\\              
   &  & $\beta_1 = 1$ & $\beta_2 = 1.4$ & $\beta_3 = 2$  \\ 
 & Average Estimates &  0.9971 & 1.3901 & 2.0006 \\ 
 & Mean Square Error & 0.0066 & 0.0117  & 0.016 \\      
 & & $\lambda_1 = 1$ & $\lambda_2 = 0.5$ & $p = 0.3$ \\ 
 & Average Estimates & 1.0279 &  0.5004  & 0.2887 \\
 & Mean Square Error & 0.0159 & 0.00022  & 0.0027 \\ \hline
Parameter Set &  & $\alpha_1 =  1$ & $\alpha_2 = 1.2$ & $\alpha_3 = 1$   \\ 
 n = 1000 & Average Estimates & 1.0457 & 1.2556  & 1.0046 \\ 
 & Mean Square Error & 0.0314 & 0.0459 & 0.0194\\              
   &  & $\beta_1 = 1$ & $\beta_2 = 1.4$ & $\beta_3 = 2$  \\ 
 & Average Estimates &  0.9988  & 1.3885   & 1.998 \\ 
 & Mean Square Error & 0.0114  & 0.0189  & 0.025 \\      
 & & $\lambda_1 = 1$ & $\lambda_2 = 0.5$ & $p = 0.3$ \\ 
 & Average Estimates & 1.00825 & 0.50148 & 0.2889 \\
 & Mean Square Error & 0.01978 & 0.00034 & 0.0027 \\ \hline
Parameter Set &  & $\alpha_1 = 0.5$ & $\alpha_2 = 0.4$ & $\alpha_3 = 0.3$ \\
 n = 1500 & Average Estimates & 0.5051 & 0.4076 & 0.2855 \\
  & Mean Square Error & 0.0012 & 0.00168 & 0.0015\\
 &  & $\beta_1 = 0.5$ & $\beta_2 = 1.5$ & $\beta_3 = 0.5$ \\
 & Average Estimates & 0.4958 & 1.4907 & 0.5391 \\
 & Mean Square Error & 0.0048 & 0.1365 & 0.0140 \\
 &  & $\lambda_1 = 2$ & $\lambda_2 = 1.5$ & $p = 0.6$ \\
 & Average Estimates & 2.0493 & 1.5128  & 0.5850 \\
  & Mean Square Error & 0.0302 & 0.0062 & 0.00495 \\ \hline
Parameter Set &  & $\alpha_1 = 0.5$ & $\alpha_2 = 0.4$ & $\alpha_3 = 0.3$ \\
 n = 1000 & Average Estimates & 0.5039 & 0.4071 & 0.2886\\
  & Mean Square Error & 0.0021 & 0.0021 & 0.0017\\
 &  & $\beta_1 = 0.5$ & $\beta_2 = 1.5$ & $\beta_3 = 0.5$ \\
 & Average Estimates & 0.5052 & 1.6028 & 0.53907\\
 & Mean Square Error & 0.0122 & 0.9205 &  0.01404\\
 &  & $\lambda_1 = 2$ & $\lambda_2 = 1.5$ & $p = 0.6$ \\
 & Average Estimates & 2.05437 & 1.5300 & 0.5847 \\
  & Mean Square Error & 0.05437 & 0.0168 & 0.00697 \\ \hline
\end{tabular}}   
\caption{The Average Estimates (AE) and Mean Square Error (MSE)}
\label{AEMSE} 
\end{table}

\section{Conclusion}

  In this paper we proposed hierarchical EM algorithm in mixture of two bivariate distributions.  We formulate the mixtures of taking higher dimensional version of Generalized Exponential distribution proposed by Kundu and Gupta \cite{GuptaKundu:1999}.  We observed that our algorithm is giving good results for large samples.  Although MSE is on higher side for small sample.  It can be a good guess for the choice of initial parameters in other optimization algorithm. 

 We can further extend this version in more generalized set-up or much larger class of distributions.  The work is on progress.

\end{document}